\documentstyle[12pt,twoside]{article}
\def\greaterthansquiggle{\raise.3ex\hbox{$>$\kern-.75em\lower1ex\hbox{$\sim$}}}
\def\lessthansquiggle{\raise.3ex\hbox{$<$\kern-.75em\lower1ex\hbox{$\sim$}}}
\newcommand{\beq}{\begin{equation}}
\newcommand{\eeq}{\end{equation}}
\newcommand{\beqa}{\begin{eqnarray}}
\newcommand{\eeqa}{\end{eqnarray}}
\newcommand{\beqan}{\begin{eqnarray*}}
\newcommand{\eeqan}{\end{eqnarray*}}
\newcommand{\ba}{\begin{array}}
\newcommand{\ea}{\end{array}}
\newcommand{\no}{\nonumber}

\newcommand{\lets}{\lessthansquiggle}

\newcommand{\ra}{\rightarrow}

\newcommand{\ve}{\varepsilon}

\newcommand{\dg}{\dagger}

\newcommand{\cL}{{\cal L}}
\newcommand{\M}{{\cal M}}

\def\nz{\ifmmode {I\hskip -3pt N} \else {\hbox {$I\hskip -3pt N$}}\fi}
\def\zz{\ifmmode {Z\hskip -4.8pt Z} \else
       {\hbox {$Z\hskip -4.8pt Z$}}\fi}
\def\qz{\ifmmode {Q\hskip -5.0pt\vrule height6.0pt depth 0pt
       \hskip 6pt} \else {\hbox
       {$Q\hskip -5.0pt\vrule height6.0pt depth 0pt\hskip 6pt$}}\fi}
\def\rz{\ifmmode {I\hskip -3pt R} \else {\hbox {$I\hskip -3pt R$}}\fi}
\def\cz{\ifmmode {C\hskip -4.8pt\vrule height5.8pt\hskip 6.3pt} \else
       {\hbox {$C\hskip -4.8pt\vrule height5.8pt\hskip 6.3pt$}}\fi}

\def\au{{\setbox0=\hbox{\lower1.36775ex%
\hbox{''}\kern-.05em}\dp0=.36775ex\hskip0pt\box0}}
\def\ao{{}\kern-.10em\hbox{``}}

\normalsize
\begin{document}
\bibliographystyle{plain}
\begin{titlepage}
\begin{flushright}
UWThPh-1994-51\\
CFIF--IST--4/94 \\
\today
\end{flushright}
\vspace{2cm}
\begin{center}
{\Large \bf  $D^0 - \bar D^0$ Mixing in the Presence of Isosinglet
Quarks}\\[30pt]
G.C. BRANCO*, P.A. PARADA**  \\
Departamento de F\'\i sica and CFIF-IST\\
Instituto Superior T\'ecnico \\
Avenida Rovisco Pais, 1096 Lisboa -- Codex, Portugal \\[5pt]
and \\[5pt]
M.N. REBELO*** \\
Institut f\"ur Theoretische Physik, \\
Universit\"at Wien, Boltzmanngasse 5, \\
A-1090 Wien, Austria \\

\vfill
{\bf Abstract} \\
\end{center}

We analyse $\Delta C = 2$ transitions in the framework of a minimal
extension of the Standard Model where either a $Q = 2/3$ or a
$Q = - 1/3$ isosinglet quark
is added to the standard quark spectrum. In the case of a $Q=2/3$
isosinglet quark,
it is shown that there is
a significant region of parameter space where $D^0 - \bar D^0$ mixing
is sufficiently enhanced to be observed at the next round of
experiments. On the contrary, in the case of a $Q=-1/3$ isosinglet
quark, it is pointed out that obtaining a substancial enhancement
of $D^0-\bar D^0$ mixing, while complying with the experimental
constraints on rare kaon decays, requires a contrived choice of
parameters.

\vfill
\begin{enumerate}
\item[*)] Work  supported in part by Science project No
SCI--CT91--0729 and EC contract No CHRX--CT93--0132.
\item[**)] Presently supported by JNICT--Programa Ciencia Grant
BD/1504/91--RM.
\item[***)] Presently supported by Fonds zur F\"orderung der wissenschaftlichen
Forschung, Project No. P8955--PHY.
\end{enumerate}
\end{titlepage}

\section{Introduction}
Flavour--changing processes leading to $K^0 - \bar K^0$ and
$B^0 - \bar B^0$ mixings have played an important r\^ole in testing the
Standard Model (SM) and in constraining some of the physics beyond the
SM. Indeed the $K_L - K_S$ mass difference was used to predict the
approximate value of the charm quark mass [1] while the experimental
value of $B_d - \bar B_d$ mixing [2] provided the first hint that the
top quark is much heavier than anticipated. Regarding $D^0 - \bar D^0$
mixing, its main interest stems from the fact that it provides a
sensitive probe of physics beyond the SM. Within the SM the
short--distance contributions to $D^0 - \bar D^0$ mixing arise from
box--diagrams and are small, $\Delta M_D^{\rm box} (SM) \approx
10^{-17}$~GeV. Initially, it was argued that the mass difference in
the $D^0 - \bar D^0$ sector was mainly a long distance effect and a
discussion of both short and long distance effects led to the
estimate [3,4] that the long distance contribution is about two orders
of magnitude larger than the box diagram contribution. A subsequent
study suggested [5] that there are significant cancellations among the
dispersive channels, leading to the prediction that the long distance
contribution is smaller than previously estimated. This conjecture was
recently confirmed by an analysis of $D^0 - \bar D^0$ mixing in
heavy quark effective field theory (HQEFT), including leading order
QCD corrections, which led to the result [6]
$(\Delta M_D)_{HQEFT} \approx (0.9 - 3.5) \cdot 10^{-17}$~GeV. These
results imply that the observation of $D^0 - \bar D^0$ mixing at the
next round of experiments at a tau/charm factory, at Fermilab and at
CESR, would provide a clear indication of physics beyond the SM, since
the present experimental bound is still four orders of magnitude above
the SM prediction and the expected sensitivity of these experiments
would easily allow for an improvement of two orders of magnitude. The
simplest mechanism to obtain significant new contributions to
$D^0 - \bar D^0$ mixing consists of having either scalar or gauge
flavour--changing neutral currents (FCNC). The importance of Higgs
couplings to FCNC and their impact on $D^0 - \bar D^0$ mixing have
been recently emphasized by Hall and Weinberg [7]. Models with isosinglet
quarks provide the simplest framework where $Z$ flavour--changing
neutral currents (ZFCNC) arise at tree level, together with deviations
from unitarity of the CKM matrix.

In this paper, we will analyse $D^0 - \bar D^0$ mixing within a minimal
extension of the SM with either a $Q = 2/3$ or a $Q = - 1/3$ isosinglet
quark. The description of the model is provided in section 2. In
section 3 we analyse in detail the contribution to $D^0 - \bar D^0$
mixing
and point out that in models with a $Q=2/3$ isosinglet quark and
for plausible values of the parameters, the mixing can be sufficiently
enhanced to be observed at the next round of experiments. This
is to be contrasted with the case of models with a $Q=-1/3$
isosinglet quark, where we show that achieving a significant
enhancement of $D^0 - \bar D^0$ mixing while satisfying
the experimental constraints on rare kaon processes requires
fine-tuning of parameters.
We also consider, in section 4, the possibility of having an
enhancement of single top production in $e^+ e^-$, having in mind a
search for this process in LEP~II. However, we show that only for a
somewhat contrived choice of parameters, is this search feasible at
LEP~II. In section 5, we present our conclusions.

\section{Model with $Q = 2/3$ Isosinglet Quark}
\renewcommand{\theequation}{\arabic{section}.\arabic{equation}}
\setcounter{equation}{0}

In order to settle the notation and provide a framework to discuss
ZFCNC, we briefly describe a minimal model where ZFCNC arise in the
up quark sector. We will consider a minimal extension of the SM which
consists of adding a charge (2/3) quark $T$ whose left--handed and
right--handed components are both singlets under $SU(2)$. One may give
appropriate mass to all quarks using only one Higgs doublet as in the
SM. In this case, apart from the SM Yukawa couplings one has the
$SU(2) \times U(1)$ invariant mass terms $M_j \bar T_L u_{R_j}$.
A slightly more complicated Higgs structure is required in order to
obtain spontaneous CP breaking and a possible solution to the strong
CP problem. In that case, it has been shown [8] that the minimal
Higgs structure consists of introducing a complex $SU(2) \times U(1)$
singlet, together with the SM Higgs doublet. Our analysis does not
depend on whether one introduces only a Higgs doublet or a Higgs
doublet and a singlet.

Without loss of generality, one may choose a weak--basis where the
$3 \times 3$ down--quark mass matrix is diagonal, real and the
$4 \times 4$ up--quark mass matrix has the form:
\beq
\M_u = \left[ \ba{cc} G & J \\ 0 & M \ea \right]
\eeq
where $G$ is a $(3 \times 3)$ matrix, while $J$ is a $(3 \times 1)$
column matrix. $M$ corresponds to a good approximation to the mass
of the isosinglet quark $T$. If we denote by $m$ the mass scale of
$G$, $J$, then it is natural to assume $M \gg m$, since $G$, $J$ are
$\Delta I = 1/2$ mass terms while $M$ is a $\Delta I = 0$ mass term.
The weak--eigenstates $(u^0_i,T^0)$ and mass--eigenstates $(u_i,T)$
are related through the unitary transformation:
\beq
\left( \ba{c} u^0 \\ T^0 \ea \right)_L =
W \left[ \ba{c} u_i \\ T \ea \right]_L .
\eeq
It is convenient to write the $4 \times 4$ unitary matrix $W$ as:
\beq
W = \left[ \ba{cc} K & R \\ S & X \ea \right]
\eeq
where $(K\;R)^\dg$ is the $(4 \times 3)$ CKM matrix, while $K^\dg$
corresponds to its $(3 \times 3)$ block connecting standard quarks.

The charged--current interactions can then be written as:
\beq
\frac{g}{\sqrt{2}} \left[\bar u_L K^\dg \gamma_\mu d_L + \bar T_L R^\dg
\gamma_\mu d_L\right] W^\mu + h.c.
\eeq
A nice feature of this class of models is that both deviations from
unitarity of the matrix $K^\dg$ and ZFCNC among standard quarks are
naturally suppressed
[9] by the ratio $m^2/M^2$. Indeed the unitarity of $W$ implies:
\beq
\left( V_{CKM}\right) \left( V^\dg_{CKM}\right) = 1 - S^\dg S
\eeq
where $V_{CKM} \equiv K^\dg$. An approximate diagonalization of
$\M_u \M^\dg_u$ leads to:
\beq
S \approx - \frac{J^\dg K}{M}
\eeq
and therefore $S$ is of order $m/M$.

The neutral current interactions are given by:
\beq
\cL_Z = \frac{g}{2 \cos \theta_W} \left[ z_{\alpha \beta}
\bar u_{\alpha L} \gamma^\mu u_{L \beta} - \delta_{ij} \bar d_{i_L}
 \gamma^\mu d_{j_L}
- 2 \sin^2 \theta_W J^\mu_{\rm em} \right] Z_\mu
\eeq
where $\alpha = 1 \ldots 4$, with $u_4 \equiv T$ and $u_i \equiv
(u,c,t)$. Of particular interest to us are the ZFCNC connecting
standard quarks, which are given by:
\beq
Z_{ij} = - S^*_i S_j
\eeq
where the indices $i,j$ in $Z_{ij}$ refer to $u,c,t$. A crucial
question is: for a given
value of $M$, what are the expected values of $Z_{ij}$? In order to
answer this question, we write explicitly the expressions for the
$S_i$:
\beqa
S_1 &\cong& -\frac{1}{M} \left( J^*_1 V^*_{ud} + J^*_2 V^*_{us} +
J^*_3 V^*_{ub} \right) \no \\
S_2 &\cong& -\frac{1}{M} \left( J^*_1 V^*_{cd} + J^*_2 V^*_{cs} +
J^*_3 V^*_{cb} \right)  \\
S_3 &\cong& -\frac{1}{M} \left( J^*_1 V^*_{td} + J^*_2 V^*_{ts} +
J^*_3 V^*_{tb} \right). \no
\eeqa
It is clear from Eq. (2.9) that the values of the $S_i$ crucially depend
on the values of the $J_i$. One may be tempted to consider that the $J_i$
follow the standard up quark hierarchy. We would like to point out that
this is not the case in general, since the $J_i$ are independent parameters,
unrelated to either the quark mass spectrum or the $(3 \times 3)$
$V_{CKM}$
matrix. This can be easily shown by noting that to leading order in
$m^2/M^2$, the following relation holds [9]:
\beq
K^{-1} G G^\dg K = \bar m^2
\eeq
where $\bar m^2 \equiv \mbox{diag }(m^2_u,m^2_c,m^2_t)$. Therefore, both the
spectrum of the standard up quark masses and the $(3 \times 3)$ block
of the CKM matrix, ($V_{CKM} \equiv K^\dg$), are determined, to leading
order, by the hermitian matrix $G G^\dg$. As a result, the $J_i$ should be
treated as independent parameters, unrelated to the standard quarks
mass spectrum. Thus in the next section where we analyse the
size of $D^0 - \bar D^0$ mixing, we will consider various
hypothesis for the relative size of the $J_i$.

We have described the main features of a model with a $Q = 2/3$
isosinglet quark. It is clear that entirely analogous considerations
apply to a model with a $Q = - 1/3$ isosinglet quark.

\section{$D^0 - \bar D^0$ mixing}
\renewcommand{\theequation}{\arabic{section}.\arabic{equation}}
\setcounter{equation}{0}

Within the SM, the short distance contributions to $D^0 - \bar D^0$ mixing
arise from box--diagrams and are small due essentially to the fact that
the dominant terms are those associated to the intermediate quark lines
$s$ and $d$ where the suppression factor
$(m_s^2 - m_d^2)^2/(M_W^2 m_c^2)$ appears [10]. This factor should be
compared to the corresponding factor for $K^0 - \bar K^0$ mixing which
is $m_c^2/M_W^2$ (roughly 1300 times higher for $m_s \simeq 0.2$~GeV
and $m_c \simeq 1.2$~GeV).

Next we evaluate $D^0 - \bar D^0$ mixing in models with either a
$Q = 2/3$ or a $Q = - 1/3$ isosinglet quark.

\subsection{$D^0 - \bar D^0$ mixing in a model with a $Q = 2/3$
isosinglet quark}
In this case, there are ZFCNC in the up quark sector which lead to a new
contribution to $D^0 - \bar D^0$ mixing from $Z$ exchange tree graphs
(Fig. 1). Using Eq. (2.7) one obtains for the effective $\Delta C = 2$
Lagrangian:
\beq
\cL^Z_{\rm eff} = \frac{- g^2}{2 \cos^2 \theta_W M_Z^2}{ (\frac{1}{2}
Z_{uc})^2} (\bar u_L \gamma^\mu c_L)(\bar u_L \gamma_\mu
c_L).
\eeq
{}From the relations:
\beqa
\Delta M_D &\simeq& 2 | M_{12}{}^D| \no \\
(M_{12}{}^D)_{ZFCNC} &=& - \langle D^0 | \cL^Z_{\rm eff}| \bar D^0 \rangle
\eeqa
$$
\langle D^0|(\bar u_L \gamma^\mu c_L)(\bar u_L \gamma_\mu c_L)|
\bar D^0\rangle = \frac{1}{4}\; \frac{8}{3}\; \frac{F_D^2 M_D^2}{2M_D}\;
B_D \eta
$$
one then obtains
\beq
(M_{12}{}^D)_{ZFCNC} = \frac{\sqrt{2} \; G_F \; F_D^2 \; B_D \; M_D}{6}\;
Z_{uc}^2  \eta
\eeq
where $\eta$ is a QCD correction factor expected to be of order one.

The size of the new contribution to $D^0 - \bar D^0$ mixing depends
crucially on the magnitude of $Z_{uc}$. From Eqs. (2.8), (2.5) it follows
that the $Z_{ij}$ are related to the deviations from unitarity of $V_{CKM}$.
One can then use the experimental knowledge of $V_{CKM}$ to constrain
the size of $Z_{uc}$. Explicitly, one has
\beq
|S_1|^2 = 1 - |V_{ud}|^2 - |V_{us}|^2 - |V_{ub}|^2.
\eeq
{}From the experimental values [11]:
\beqa
|V_{ud}| &=& 0.9744 \pm 0.0010 \no \\
|V_{us}| &=& 0.2205 \pm 0.0018 \no \\
|V_{ub}|/|V_{cb}| &=& 0.08 \pm 0.02  \\
|V_{cb}| &=& 0.040 \pm 0.005 \no
\eeqa
it follows that:
\beq
|S_1| \leq 7 \cdot 10^{-2} .
\eeq
A bound on $|S_2|$ may also be derived from Eq. (2.5). However, due
to the poor experimental knowledge on $|V_{cs}|$, only a loose bound
on $|S_2|$ can be obtained
\beq
|S_2| \leq 0.5.
\eeq
One may also try to obtain a limit on $|S^*_2 S_3|$ using the relation:
\beq
|S^*_2 S_3| = | V_{cd} V^*_{td} + V_{cs} V^*_{ts} + V_{cb} V^*_{tb}|.
\eeq
However, given the present knowledge of $V_{CKM}$, only a loose bound
can be obtained. If, for example, one assumes $V_{ts} \approx V_{cb}$,
it follows that
\beq
|S_2^* S_3| < 10^{-1}.
\eeq
Next we will analyse the expectations for the strength of the $Z_{ij}$,
within the present model. The magnitude of the $Z_{ij}$ crucially depends
on the assumptions one makes for the $J_i$, defined by Eq. (2.1). In order
to illustrate this dependence, we will analyse the size of the $Z_{ij}$,
for two special cases:
\begin{enumerate}
\item[i)] Non--Hierarchical $J_i$ 's

The simplest assumption one can make about the $J_i$ is that they are all
roughly of the same order of magnitude, i.e. $J_1 \sim J_2 \sim J_3
\equiv J$. Using Eqs. (2.8), (2.9) one then obtains:
\beq
|Z_{uc}| \simeq |Z_{ct}| \simeq |Z_{ut}| \simeq \frac{|J|^2}{M^2} .
\eeq
In this case, the best bound on $|J|/M$ arises from the experimental
bound on $\Delta M_D$ [11]:
\beq
(\Delta M_D)_{\rm exp} < 1.3 \cdot 10^{-13} \mbox{ GeV}.
\eeq
{}From Eqs. (3.1), (3.7), (3.8), one obtains
\beq
|J|/M < 3.3 \cdot 10^{-2}
\eeq
where we have used $B_D F_D^2 \approx 0.01$~GeV$^2$,
$M_D = 1.9$~GeV. From Eqs. (3.3), (3.7), one concludes that if one
assumes, for example, $M \sim 0.5$~TeV, then $(\Delta M_D)_{ZFCNC}$
would be two orders of magnitude larger than $(\Delta M_D)_{HQEFT}$
for $J \approx 5$~GeV. Taking into account the size of $m_t$,
it is clear that larger values of J are entirely plausible and therefore
$(\Delta M_D)_{ZFCNC}$ can be even further enhanced.

\item[ii)] Hierarchical $J_i$'s

If one assumes that the $J_i$ differ significantly from each other,
the simplest possibility is to consider that the $J_i$ follow the
standard up quark hierarchy, i.e. $J_1 \sim m_u$, $J_2 \sim m_c$,
$J_3 \sim m_t$. Using Eqs. (2.9) and putting $m_c \sim 1.2$~GeV,
$m_t \sim 174$~GeV, $V_{cb} \sim 0.05$, one obtains
\beq
|S_1| \sim \frac{1.15}{M}; \qquad
|S_2| \sim \frac{10}{M}; \qquad
|S_3| \sim \frac{174}{M} .
\eeq
If we take $M \simeq 0.5$~TeV, we obtain:
\beq
Z_{uc} = |S_1 S_2| \simeq 0.46 \cdot 10^{-3}.
\eeq
For this value of $Z_{uc}$, one obtains $(\Delta M_D)_{ZFCNC}$ about two
orders of magnitude above the value of $(\Delta M_D)_{HQEFT}$, and
therefore at the reach of the next round of experiments.
\end{enumerate}

\subsection{$D^0 - \bar D^0$ mixing in a model with a $Q = -1/3$
isosinglet quark}
If the SM is extended by adding a $Q = -1/3$ isosinglet quark $N$, there
will be ZFCNC in the down quark sector and the SM portion of the new
CKM matrix will also deviate from a unitary matrix. We may choose,
without loss of generality, a weak--basis where the $3 \times 3$ up
quark mass matrix is diagonal, and the $4 \times 4$ down quark mass
matrix $\M_d$ has the same form as that given by Eq. (2.1). If we
define the matrix $W$ in this contex as the unitary matrix which
diagonalizes $\M_d \M_d^\dg$, it can still be specified by Eq. (2.3)
where, as before, the block $S$ parametrizes the ZFCNC and the block
$R$ gives the new charged boson couplings. The charged current interactions
are now given by:
\beq
\frac{g}{\sqrt{2}} [ \bar u_L K \gamma_\mu d_L + \bar u_L R \gamma_\mu N]
W^\mu + h.c.
\eeq
In this case, the short distance contributions to the mixing in the
$D^0 - \bar D^0$ system arise through the box diagrams of Fig.~2 where
the quarks participating in the internal lines are the down
quarks $d,s,b$ and the new quark $N$, and with $R_i = V_{iN}$ in the
notation of Fig.~2. The coefficients $V_{u\alpha},V_{c\alpha}$ appearing
in the vertices of Fig.~2, correspond to the first and the second row
of the new $(3 \times 4)$ CKM matrix. Since this matrix consists of the
first three lines of a $4 \times 4$ unitary matrix, the orthogonality
relation $\sum V^*_{c\alpha} V_{u\alpha} = 0$ still holds and, as a
result, the technical details of the calculation of these box diagrams
coincide with those encountered in the framework of four standard
generation model [12]. The new aspect of the present model is that the
strength of the couplings of the extra down--quark is related to its
mass.

In the $D^0 - \bar D^0$ system, the usual zero external momentum
approximation in the evaluation of the box diagram [13] for three
generations in the SM is not justified because the masses of the
internal quarks are not always large compared to the $c$--quark mass
and a more detailed calculation has been performed by several authors
[10]. Yet, for the quark $N$, which we assume to be much heavier than
the $c$--quark, the zero external momentum approximation is good and
we have (neglecting CP violating effects)
\beq
\Delta M_D(N,N) \simeq 2|M_{12}(N,N)| \simeq \frac{G_F^2}{6\pi^2}
F^2_D B_D m_D M_W^2 |\lambda_N^2| S(x_N) \eta
\eeq
where $\lambda_N = V_{uN} V^*_{cN}$, $x_N = (m_N/m_W)^2$, $\eta$
denotes a QCD correction coefficient and $S(x_N)$ is an Inami--Lim
function:
\beq
S(x_N) = x_N \left[ \frac{1}{4} + \frac{9}{4} \frac{1}{(1 - x_N)} -
\frac{3}{2} \frac{1}{(1 - x_N)^2} \right] + \frac{3}{2}
\left[ \frac{x_N}{x_N - 1}\right]^3 \log x_N
\eeq
computed for an arbitrary value of the internal quark mass. We are
interested in the case where the contribution of the new quark $N$ is
dominant and therefore we neglect the contributions of the standard
down--type quarks. Comparing Eq. (3.15) to Eq. (2.4) one readily
concludes that the unitarity constraints on $S$ in the model with one
$Q = 2/3$ isosinglet quark coincide with the constraints on $R$ in the
model with a $Q = - 1/3$ isosinglet quark. In particular, the following
bounds have to be satisfied:
\beq
|R_1| \equiv |V_{uN}| \leq 7 \cdot 10^{-2}, \qquad
|R_2| \equiv |V_{cN}| \leq 0.5.
\eeq
In leading order one has:
\beq
R_i \cong J_i/M .
\eeq
In the estimation of the magnitude of the factor $|\lambda_N|^2 S(x_N)$,
one has to keep in mind the following points. On the one hand, given
the value of $m_b$, a value of $m_N$ of order e.g. 200~GeV, would still
be compatible with the approximation used in the diagonalization of
$\M_d \M_d^\dg$, since one would have $m_b^2/m_N^2 \ll 1$. On the other
hand, one has to take into account the experimental constraints on
ZFCNC, which in the present model, with a $Q = - 1/3$ isosinglet quark,
appear in the down--quark sector. The most stringent constraints arise from
$K_L \ra \mu^+ \mu^-$ and the size of the CP violating parameter $\ve$,
which lead to the following bounds [14]:
\beq
|\mbox{Re }(Z_{ds})| \leq 2.6 \cdot 10^{-5}, \qquad
|\mbox{Im }(Z_{ds})^2| \leq 9.2 \cdot 10^{-10}.
\eeq
We should keep in mind that in the model with a $Q = - 1/3$ isosinglet
quark, one has relations entirely analogous to Eqs. (2.6), (2.8). In
particular,
\beq
Z_{ds} = - S_1^* S_2
\eeq
with the $S_i$ given by Eq. (2.6). If we assume $J_1 \sim J_2 \sim J_3
\equiv J$, the bounds of Eq. (3.20) lead to:
\beq
J/M < 5 \cdot 10^{-3}
\eeq
which in turn implies:
\beq
\lambda_N \; \lets \; 2.5 \cdot 10^{-5}.
\eeq
For this value of $\lambda_N$, the contribution of the isosinglet quark
to $\Delta M_D$ would be negligible. At this point, one may ask whether
it is possible at all to obtain a significant contribution to
$\Delta M_D$ in models with a $Q = - 1/3$ isosinglet quark. From Eqs. (3.19),
(3.20), (3.21) it is clear that obtaining a sizable contribution to
$\Delta M_D$ while conforming to the bounds of Eq. (3.20) requires a
strong cancellation among the various terms contributing to $S_1$.
Assuming that this cancellation occurs, then for $J_1 \cong 3$~GeV,
$J_2 \cong 15$~GeV, $M_N \cong 200$~GeV, one obtains:
\beq
\lambda_N \cong 1.1 \cdot 10^{-3}
\eeq
which leads to:
\beq
\Delta M_D(N,N) \sim  10^{-15}\mbox{ GeV}
\eeq
which is roughly two orders of magnitude larger than
$(\Delta M_D)_{HQEFT}$. However, it should be emphasized that in the
model with a $Q = - 1/3$ isosinglet quark, a large value for
$\Delta M_D(N,N)$, consistent with the bound on $Z_{ds}$, can only be
obtained by assuming a contrived choice of parameters leading to the
suppression of $S_1$. This is to be contrasted with the situation one
encountered in the model with a $Q = 2/3$ isosinglet quark, where a
large contribution to $\Delta M_D$ is obtained without assuming any
fine--tuning of parameters.

\section{Single Top Production at LEP~II}
\renewcommand{\theequation}{\arabic{section}.\arabic{equation}}
\setcounter{equation}{0}

In the SM, the cross section for single top quark production through
the reaction $e^+ e^- \ra t \bar b e \bar \nu_e$ is too small [15]
for the detection of single top quark to be feasible at LEP~II energy
and planned luminosity. In the presence of ZFCNC in the
$Q = 2/3$ quark sector, the lowest order single top quark production
process $e^+ e^- \ra t \bar c(t \bar u)$ is possible, thus providing a
potential probe of ZFCNC involving the top quark. The possibility of
searching for $t$--flavour violating neutral currents via single top
quark production at $e^+ e^-$ colliders was raised about ten years ago by
Hikasa [16]. However, the top quark mass is now known to be much larger
than what was generally expected at that time. The overall strength
and the structure of the $t$--flavour changing neutral current
interactions can be parametrized in the following way:
\beq
\cL = i \frac{g}{2\cos \theta_W} z_{tq} \bar q \gamma_\mu (V_q - A_q \gamma_5)t
Z^\mu + h.c.
\eeq
where $q = u,c$. The cross section for the $Z$ mediated single top
production process is then given by:
\beq
\sigma_Z(e^+ e^- \ra t \bar q) = \frac{G_F^2}{8 \pi} (1 - 4 \sin^2 \theta_W
+ 8 \sin^4 \theta_W) \frac{M_Z^4(s - m_t^2)^2(2s + m_t^2)}{s^2(s - M_Z^2)^2}
(V_q^2 + A_q^2) |z_{tq}|^2
\eeq
where, in view of the very high value of the top quark mass, we have safely
neglected the $Z$--width and $m_q \equiv m_c,m_u$. The result of
Eq. (4.2) is model independent. We can particularize to the present model
where ZFCNC arise as a result of the mixing of a $Q = 2/3$ isosinglet quark
with
the standard quarks, by setting $A_q = V_q = 1/2$. Having in mind LEP~II,
we have plotted in Fig.~3, $y=\sigma/|z_{tq}|^2$ against $m_t$, for
$\sqrt{s} = 175$, 190~GeV. If we consider an integrated luminosity of
500~pb$^{-1}$, then the values of $|z_{tq}|$ corresponding to five events
are shown in Fig.~4. It is clear that only for relatively large values of
$Z_{tq}$ $(Z_{tq} \sim 0.1)$ is the detection of single top at LEP~II
possible. We turn now to the question whether such a value of $Z_{tq}$ is
possible in the framework of the model with a $Q = 2/3$ isosinglet quark.
{}From Eqs. (2.8), (2.9) it is clear that one may have $Z_{tc} \sim (0.1)$,
by choosing, e.g. $J_2 \approx J_3 \approx m_t$ and $m_Q \approx 0.5$~TeV.
However, taking into account $Z_{uc} = |S_1 S_2|$, it is clear from
Eqs. (2.9) that these value of $J_2$, $J_3$ would lead to a value of
$\Delta M_D$ which would violate the experimental bound of Eq. (3.11),
unless  there is a cancellation in the various terms contributing to
$S_1$. Therefore, in the model with a $Q = 2/3$ isosinglet quark, only
for a contrived choice of parameters can one obtain $Z_{tc}$ sufficiently
large to lead to single top production at LEP~II and at the same time
have $Z_{uc}$ sufficiently small to conform to the experimental bound
on $\Delta M_D$.

\section{Conclusions}
\renewcommand{\theequation}{\arabic{section}.\arabic{equation}}
\setcounter{equation}{0}
We have investigated the impact of ZFCNC on $D^0 - \bar D^0$ mixing and on
single top production, in the framework of an extension of the SM where
isosinglet quarks are introduced. It was shown that in models with a
$Q = 2/3$ isosinglet quark, with a mass of the order of one TeV, the new
contribution to $\Delta M_D$ arising from ZFCNC can be more than two
orders of magnitude larger than the SM contribution evaluated within the
framework of HQEFT, and therefore at the reach of the next round of
experiments. This enhancement of $D^0 - \bar D^0$ mixing is obtained for
generic values of the parameters and does not involve any fine--tuning.
On the contrary, in models with a $Q = - 1/3$ isosinglet quark, due to
the constraints arising from rare kaon decays and the value of $\ve$,
only for special choices of parameters leading to the
necessary cancellations can one still
obtain a sizable contribution to $D^0 - \bar D^0$ mixing. At this
point, it is worth recalling that the study of CP asymmetries in $B^0$
decays provides the best way to test models with a $Q = - 1/3$ isosinglet
quark. Indeed, it has been shown [17] that even if the contribution of
ZFCNC to $B^0 - \bar B^0$ mixing is only at the level of 20 per cent, the
predictions for CP asymmetries in $B^0$ decays may differ drastically from
those of the SM.

As far as single top production is concerned, we have seen that there is
a new contribution to single--top production in $e^+ e^-$ reactions, due
to $t$--flavour violating neutral currents, which arise in models with
a $Q = 2/3$ isosinglet quark. However, it was shown that only for a
somewhat contrived choice of parameters, is the strength of ZFCNC
involving the top quark sufficiently large to lead to the detection
of single top at LEP II energy and luminosity.

\section*{Figure Captions}
\begin{description}
\item[Fig. 1] Tree diagram contribution to the $\Delta C = 2$ transition
in a model with a $Q = 2/3$ isosinglet quark.
\item[Fig. 2] One of the box diagrams that induce a $\Delta C = 2$
transition in a model with a $Q = -1/3$ isosinglet quark.
\item[Fig. 3] The value of $y = \sigma/|z_{tq}|^2$ (in pb) as a function
of the top quark mass for C.M. energies of 175~GeV (lower curve) and
190~GeV (upper curve).
\item[Fig. 4] The value of $|z_{tq}|_5$ corresponding to 5 events at the
predicted LEP~II integrated Luminosity of 500~pb$^{-1}$, as a function
of the top quark mass, for C.M. energies at 175~GeV (upper curve) and
190~GeV (lower curve).
\end{description}

\end{document}